# Proton induced reactions on $^{118}$Sn target at energies up to 18 MeV

G. H. Hovhannisyan[1,a], N.S. Gharibyan[1], T. M. Bakhshiyan[1], A. R. Balabekyan[1], S.V. Gaginyan[1], G.V. Martirosyan[1], A. Manukyan[2], R.K. Dallakyan[2], A. Aprahamian[3]

[1] Yerevan State University, 0025 Yerevan, Armenia
[2] A. Alikhanyan National Science Laboratory, 0036 Yerevan, Armenia
[3] University of Notre Dame, Notre Dame, Indiana 46556, USA
[a] e-mail: hov_gohar@ysu.am (corresponding author)

**Abstract**

Proton-induced reactions on enriched $^{118}$Sn up to 18 MeV have been investigated. Using the stacked-foil activation technique, the excitation functions of the reactions $^{118}$Sn(p,n)$^{118}$Sb, $^{118}$Sn(p,2n)$^{117}$Sb, $^{118}$Sn(p,α)$^{115m}$In, and $^{118}$Sn(p,x)$^{117m}$Sn were measured. The available experimental data show good agreement with our measurements. The cross sections for the $^{118}$Sn(p,x)$^{117m}$Sn and $^{118}$Sn(p,α)$^{115m}$In reactions are reported for the first time.

The measured cross sections were compared not only with previously published experimental results, but also with theoretical predictions from the TENDL-2023 (TALYS-based evaluated nuclear data library), TENDL-2025 and JENDL-5 (Japanese Evaluated Nuclear Data Library) libraries. Discrepancies between experimental and theoretical data were observed for reactions involving composite-particle emission, such as alpha particles and deuterons. These differences suggest that while current models adequately describe simple two-nucleon emission channels, further refinements are needed, particularly for modeling composite-particle emission at lower proton energies.

## 1 Introduction

Nuclear technologies are widely used in modern society, from research and reactor operation to managing spent fuel and producing medical isotopes. The development of these technologies has driven a growing need for accurate modeling and simulation tools capable of describing various nuclear processes. Such tools rely heavily on evaluated nuclear data libraries, in particular nuclear reaction cross-section databases.

While extensive experimental data exist for neutron-induced reactions, information on charged-particle–induced reactions remain limited. In the absence of measurements, theoretical model calculations are commonly used to provide cross-section data. Among the most widely employed sources are the TALYS-based TENDL library [1] and the Japanese evaluated nuclear data library JENDL-5 [2].

Proton-induced reactions on enriched tin isotopes are of particular interest due to their relevance for both nuclear fuel cycle studies and medical isotope production. Tin isotopes are produced as fission products of uranium and are present in spent nuclear fuel. Reactions leading to the transmutation of tin isotopes, especially $^{126}$Sn, are important for nuclear waste management, while medically relevant isotopes, such as $^{117}$Sb, $^{119}$Sb, and $^{124}$Sb can be generated via proton-induced reactions on stable tin isotopes.

Experimental cross-section data for proton-induced reactions on enriched tin targets are available; however, they are largely limited to proton energies below 9 MeV [3–22], leaving the region around the excitation function maxima insufficiently studied. In our previous work, we reported cross sections for reactions on $^{114}$Sn and $^{120}$Sn targets at proton energies up to 18 MeV, including $^{114}$Sn(p,α)$^{111(m+g)}$In, $^{114}$Sn(p,x)$^{113(m+g)}$Sn, and $^{120}$Sn(p,α)$^{117(m+g)}$In [23,24]. In the present work, we extend these studies by providing new experimental cross-section data for proton-induced reactions on $^{118}$Sn, namely $^{118}$Sn(p,n)$^{118m}$Sb, $^{118}$Sn(p,2n)$^{117}$Sb, $^{118}$Sn(p,x)$^{117m}$Sn and $^{118}$Sn(p,α)$^{115m}$In. These data offer a valuable benchmark for assessing the predictive capabilities of nuclear reaction codes and evaluated data libraries for proton-induced reactions on tin isotopes.

## 2 Experimental details

A stack of foils enriched in $^{118}$Sn (98%) was irradiated with an 18 MeV proton beam obtained from the compact medical cyclotron IBA Cyclone 18/18 located in Yerevan, Armenia. The stack consisted of 12 blocks of $^{nat}$Cu–$^{118}$Sn layers. For target set preparation, enriched tin foils (100 μm thick) were rolled and cut into disks, resulting in targets with thicknesses ranging from 23 to 54 μm, with a thickness variation within each foil of no more than 2 μm. Copper foils were used both for beam energy degradation and for monitoring beam parameters. We used 20 μm-thick copper foils as a monitor for proton flux and energy determination, as well as a degrader to reduce the proton beam energy. All foils were tightly pressed against each other and placed in an aluminum holder. The aluminum holder was placed in the cooled target module. The targets were irradiated with the proton beam for 6 min 43 sec at a current of 1 μA. The diameter of the collimated proton-beam matched with the diameter of the targets and was equal to 1.1 cm.

After irradiation, the foils in the stack were separated, and the γ-spectra of each target were measured using a high-purity germanium (HPGe) detector with an energy resolution of 1.66 keV FWHM at the 1332.5 keV peak of $^{60}$Co, and 0.618 keV FWHM at the 122 keV peak of $^{57}$Co. The detector efficiency was estimated using standard γ-sources – $^{152}$Eu, $^{133}$Ba, $^{137}$Cs, $^{60}$Co, and $^{22}$Na – with known activities supplied by Spectrum Techniques (USA), covering the entire energy range of the studied γ-rays. Fig. 1 shows the efficiency calibration curves for the two distances at which measurements were taken. The uncertainties of the experimental points were calculated using error propagation by quadrature, where the square root of the sum of the squares of the individual relative uncertainties was taken, including both statistical measurement errors and uncertainties in the monitor reaction cross-sections. The fitting was performed considering these uncertainties based on the function

$$\varepsilon(E) = e^{(a_0 + a_1 \ln E + a_2 (\ln E)^2 + a_3 (\ln E)^3 + a_4 (\ln E)^4 + a_5 (\ln E)^5)} \qquad (1).$$

The mean differences between the efficiency values and the fitted curve are 2.8% at 5 cm and 6.2% at 20 cm.

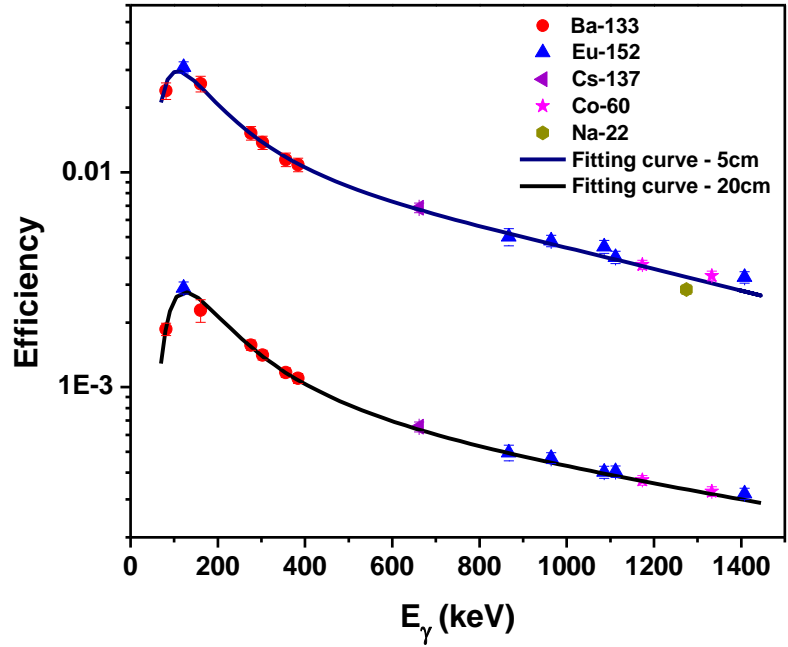

**Fig. 1.** Gamma-ray detection efficiency measured at a 5 and 20 cm source-detector distance. The error bars of the experimental energy estimation include the statistical uncertainty of the measurements and the uncertainty of the recommended monitor reaction cross sections. The fitting was performed based on the function $\varepsilon(E) = e^{(a_0 + a_1 \ln E + a_2 (\ln E)^2 + a_3 (\ln E)^3 + a_4 (\ln E)^4 + a_5 (\ln E)^5)}$.

γ-spectra of the irradiated target are measured periodically during several days, with the first measurement done about 40 min after the end of irradiation to allow short-lived isotope cross-section measurements. A typical γ-spectrum of the irradiated target is plotted in Figure 2.

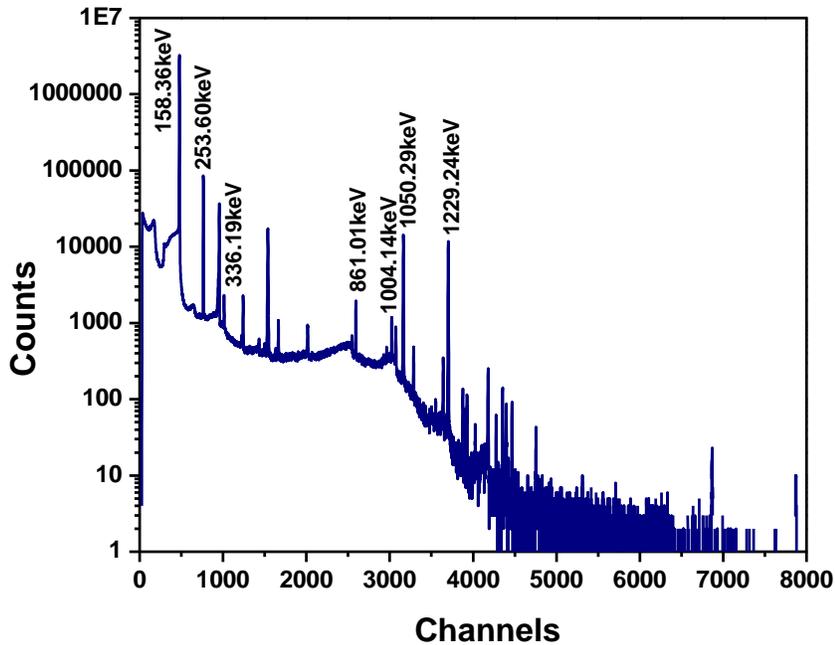

**Fig. 2.** A typical γ-spectrum of the irradiated target.

# 3 Data Evaluation

## 3.1 Determination of Energy Loss

The use of a stacked-foil arrangement consisting of natural copper $^{nat}$Cu and $^{118}$Sn foils provides the advantage of obtaining the excitation functions of the nuclear reactions of interest within a single irradiation. The basic principle is that, as the proton beam passes through the stack, its energy gradually decreases. Consequently, the activation produced in different targets corresponds to different mean beam energies.

In this configuration, the copper foils placed between the tin foils serve multiple purposes. They act both as monitor foils and as energy degraders, and they also help to capture high-energy reaction products that may escape from the tin targets. Kinematic estimates indicate that the fraction of escaping product nuclei does not exceed 0.11%, allowing this effect to be safely neglected.

The determination of the proton energy in each layer of the stack is usually [22-26] performed using simulation codes such as the SRIM [28] code. In many studies, the average proton energy in each layer is defined as the mean value between the incoming energy $E_{in}$ and the outgoing energy $E_{out}$ from that layer:

$$E_{av} = \frac{E_{in}+E_{out}}{2}. \tag{2}$$

The corresponding energy uncertainty is commonly estimated as

$$\Delta E = \frac{E_{in}-E_{out}}{2} \tag{3}$$

Fisichella et al. [27] proposed an alternative approach for determining the effective energy for each layer. In particular, the authors considered fusion reactions occurring at energies below the Coulomb barrier, where the reaction cross section, $\sigma(E)$, exhibits a strong and rapid energy dependence. Even if the incident beam initially has a relatively narrow energy spread, this spread gradually increases as the beam traverses successive foils in the stack. Under these conditions, the commonly used approximation (2) may not be fully accurate, as it does not account for the fact that each reaction energy within the target contributes with a different weight. Therefore, Fisichella et al. proposed, instead of using the average energy $E_{av}$, to determine an effective reaction energy, $E_{eff}$, which properly accounts for the distribution of incident particles over energy, as defined by equation (4).

$$E_{eff} = \frac{\int_0^\infty E\sigma(E)D(E,t_0)dE}{\int_0^\infty \sigma(E)D(E,t_0)dE} \tag{4}$$

where $D(E, t_0)$ is the probability that a beam particle will have energy $E$ inside the considered target layer of thickness $t_0$, $\sigma(E)$ is the energy-dependent reaction cross section [27].

In this equation, the influence of foil-thickness non-uniformity on the beam energy distribution in stacked targets can also be taken into account [27]. In that case, $D(E, t_0)$ includes an additional term representing the thickness probability distribution function. In our case, however, it was not possible to measure the thickness probability distribution directly. The foils are not perfectly uniform in thickness, but the non-uniformity is small (less than 6%), and this effect has been included in the uncertainties of the determined energies.

We performed calculations using the SRIM code to obtain the proton energy distributions in each layer of the stack (see Fig. 3). The distributions are generally close to Gaussian, with the exception of the first two targets, where the distributions deviate from a Gaussian shape and approach a rectangular form.

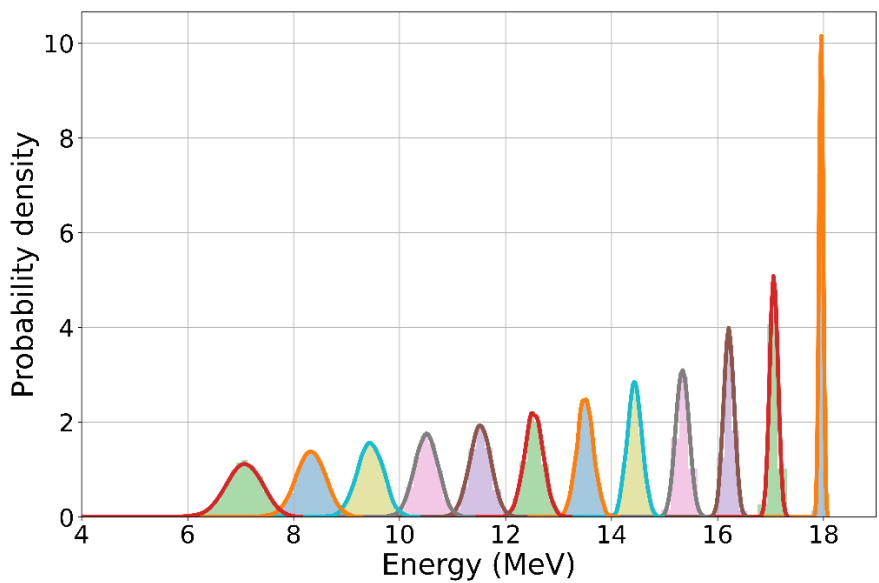

**Fig. 3.** The energy distributions of protons within the tin layer of the stacked target.

Based on SRIM calculations, the values of the average energy $E_{av}$ were determined according to equation (2) for each layer.

The effective energy $E_{eff}$ for each layer was obtained by considering the proton energy distribution within the foil. It was calculated according to equation (4), where $D(E, t_0)$ is the normalized proton energy distribution obtained from the SRIM simulation, and the cross section $\sigma(E)$ was parameterized using an exponentially modified Gaussian (EMG) function fitted to the IAEA-recommended data [29] for the $^{nat}Cu(p,x)^{63}Zn$ monitor reaction. The integrals were evaluated numerically using the trapezoidal rule over the energy interval corresponding to the simulated proton energies.

The comparison indicates that the values of energy $E_{av}$ and $E_{eff}$ obtained using the two approaches differ only slightly, by about 0.3–3% (see Fig. 4).

To verify the obtained results, we performed an experimental evaluation of the initial beam energy. The method is based on the ratio of cross sections of monitor reactions. Since the cross sections of monitor reactions are measured with high accuracy, and the ratio of the cross sections for the reactions $^{nat}Cu(p,x)^{62}Zn$ and $^{nat}Cu(p,x)^{63}Zn$ is highly sensitive to changes in the incident proton energy, this ratio can be used for beam energy determination.

The cross-section values for the above-mentioned reactions recommended by the IAEA [29] were compared with the ratio measured in our experiment. The experimental ratio was obtained from the 20 μm thick monitor foil placed before the first $^{118}Sn$ target layer. The experimental ratio of the cross sections for the monitor reactions was calculated using equation (5):

$$\frac{\sigma_{62_{Zn}}}{\sigma_{63_{Zn}}} = \frac{A_{62_{Zn}}\left(1-e^{-\lambda_{63_{Zn}}t_1}\right)}{A_{63_{Zn}}\left(1-e^{-\lambda_{62_{Zn}}t_1}\right)} \tag{5}$$

where $A_{62_{Zn}}$ and $A_{63_{Zn}}$ are the activities at the end of bombardment (EOB), $\lambda_{63_{Zn}}$ and $\lambda_{62_{Zn}}$ are the decay constants of $^{62}$Zn and $^{63}$Zn, respectively, and $t_1$ is the irradiation time [30].

From these measurements, the incident beam energy was estimated to be (18.2 ± 0.2) MeV.

To verify the SRIM results across the entire stack, the same method was applied to three additional monitor foils. However, for one of the monitor foils, where the proton energy was below the reaction threshold for the $^{65}$Cu(p,2n)$^{63}$Zn reaction, the ratio $^{nat}$Cu(p,x)$^{63}$Zn / $^{nat}$Cu (p,x)$^{65}$Zn was used instead. Experimental energy determination was possible only for four monitor targets due to the relatively short half-life of $^{63}$Zn.

Figure 4 shows the dependence of the proton energy on the target number in the stack. A good agreement between the SRIM calculations and the experimentally measured energies is observed (see Fig. 4). Figure 5, in turn, presents the dependence of the recommended [29] cross-section ratios on the proton energy, with the results of our measurements superimposed on the curves.

Fisichella et al. [27] noted that ambiguities in the determination of the beam energy may lead to misinterpretation of the derived excitation function, particularly in the exponential region of the cross section below the Coulomb barrier. In the present work, however, the beam energies are above the barrier and the corresponding variations of the reaction cross sections are relatively moderate. Therefore, the uncertainty in the adopted beam energies has little influence on the physical conclusions of this study.

Another important aspect is the effect of target inhomogeneity on the energy variation along the stack. In Ref. [27], very thin targets of about 0.7 μm were used. The target was produced by evaporating the enriched $^{120}$Sn isotope onto a rolled $^{93}$Nb foil substrate, and the thickness inhomogeneity was estimated to be as high as 150%. In the present work, foils with thicknesses of several tens of micrometers and thickness variations of up to about 6% were used. Such a level of inhomogeneity has only a minor influence on the results.

Overall, it can be concluded that the use of the simple expression for $E_{av}$ is justified for stacks consisting of nearly homogeneous foils with thicknesses of several tens of micrometers in the considered energy range.

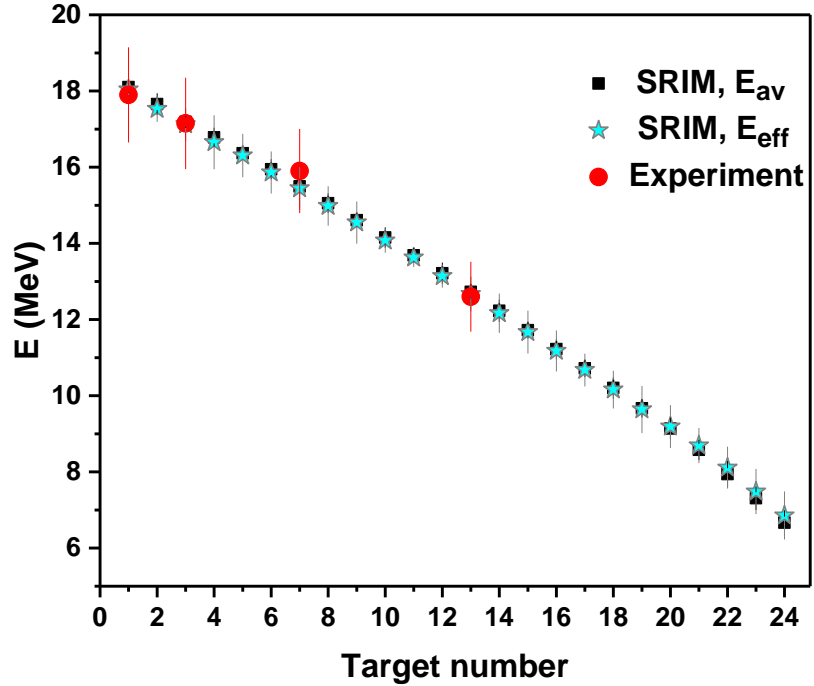

**Fig. 4.** Dependence of the proton energy on the target number in the stack. Target number 1 corresponds to the first monitor foil; the subsequent targets alternate between tin and copper foils. Red dots represent the experimentally determined energies, black squares correspond to the SRIM-based calculated average energies $E_{av}$, and blue stars indicate the SRIM-based effective energies $E_{eff}$ (see the text). The error bars of the experimental energy estimates include the statistical uncertainties of the measurements and the uncertainties of the recommended monitor reaction cross sections. For the SRIM calculations, the uncertainty of $E_{av}$ is defined as $(E_{in} - E_{out})/2$, while the uncertainty of $E_{eff}$ corresponds to the standard deviation of the energy distribution obtained from the SRIM energy spectra.

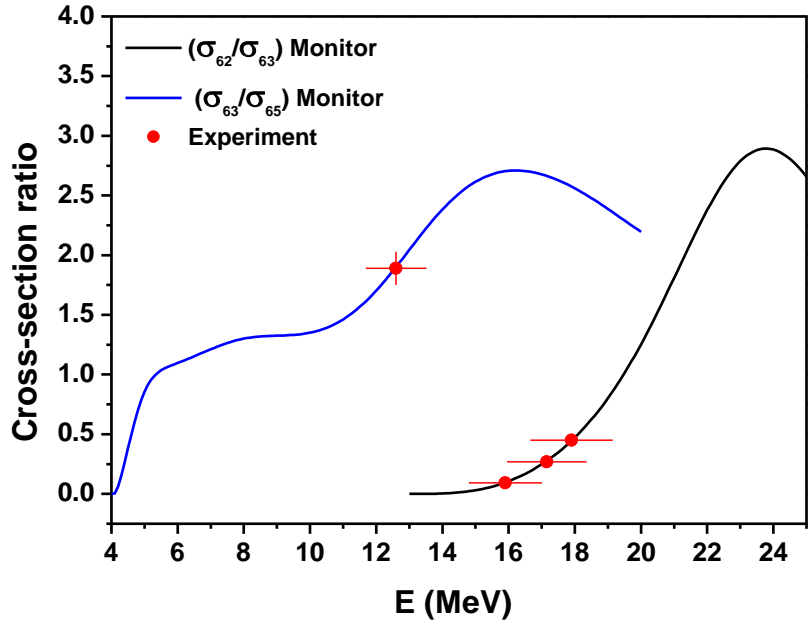

**Fig.5**. Dependence of the cross-section ratios ($\sigma_{62}/\sigma_{63}$) and ($\sigma_{63}/\sigma_{65}$) on proton energy for the recommended monitor reaction values [29] and the experimentally measured cross-section ratios. The error bars in the experimental cross-section ratios reflect the statistical uncertainties associated with the measurements.

### 3.2 Determination of cross sections

Cross sections for the reactions $^{118}$Sn(p,n)$^{118}$Sb, $^{118}$Sn(p,2n)$^{117}$Sb, $^{118}$Sn(p,x)$^{117m}$Sn, and $^{118}$Sn(p,α)$^{115m}$In were measured. The cross sections σ were determined according to Equation (5)

$$\sigma = \frac{A_{obs}\lambda\frac{t_{3r}}{t_{3l}}}{\Phi N_{nucl}\varepsilon I_\gamma (1-e^{-\lambda t_1})e^{-\lambda t_2}(1-e^{-\lambda t_{3r}})} \quad (5).$$

Here, $A_{obs}$ is the observed number of γ-rays under the photo-peak, $\lambda$ is the decay constant, $\varepsilon$ is the detector efficiency, $N_{nucl}$ is the number of target nuclei per area, Φ is the proton flux, $I_\gamma$ is the intensity of the product gamma line, $t_1$ is the irradiation time, $t_2$ is the time between the end of the bombardment and the beginning of the measurement, $t_{3r}$ is the measurement real time, and $t_{3l}$ is the measurement live time [23, 24].

The proton flux, Φ, was determined from the first copper foil of the stack using the inverse of Eq. (5) and the recommended cross sections for the monitor reactions $^{nat}$Cu(p,x)$^{62,63,65}$Zn [29]. The proton flux was assumed to be constant throughout the entire target stack.

The types of reactions and the corresponding residual nuclei, together with their characteristics, are presented in Table 1.

**Table 1.** Reaction channels, residual nuclei, Q-value, reaction threshold energies, and nuclear characteristics of the products. Threshold energies were calculated using the NNDC Q-value Calculator [31], while decay data were adopted from NuDat 3.0 [32].

| Reaction channel | Residual nucleus | Q value (MeV) | $E_{th}$ (MeV) | $T_{1/2}$ of residual nucleus | $E_\gamma$ (keV) | $I_\gamma$ (%) |
|---|---|---|---|---|---|---|
| $^{118}$Sn(p,n) | $^{118m}$Sb | -4.44 | 4.48 | 5 h | 253.678<br>1050.65<br>1229.69 | 99<br>97<br>100 |
| $^{118}$Sn(p,2n) | $^{117}$Sb | -11.87 | 11.97 | 2.8 h | 861.35<br>1004.51 | 0.31<br>0.21 |
| $^{118}$Sn(p,α) | $^{115m}$In | 2.75 | 0 | 4.486 h | 336.24 | 45.9 |
| $^{118}$Sn(p,pn)<br>$^{118}$Sn(p,d) | $^{117m}$Sn | -9.33<br>-7.10 | 9.41<br>7.16 | 13.6 d | 158.56 | 86 |

**4 Results and Discussion**

**4.1 $^{118}$Sb production**

The isotope $^{118}$Sb is produced in the (p,n) reaction in both the ground state ($J^\pi = 1^+$, $T_{1/2} = 3.6$ min) and the metastable state ($J^\pi = 8^-$, $T_{1/2} = 5.0$ h). Both states decay independently to stable $^{118}$Sn. Since the short-lived ground state cannot be detected using the off-line activation technique, only the independent production cross section of $^{118m}$Sb was reported. The cross section was determined using the interference-free γ-lines listed in Table 1.

**4.2. $^{117}$Sb production**

$^{117}$Sb is produced in the (p,2n) reaction. The independent production cross section of the $^{117}$Sb ($J^\pi = 5/2^+$, $T_{1/2} = 2.8$ h) was determined using the interference-free γ-lines listed in Table 1. Since the reaction threshold is 11.97 MeV, cross sections are reported only for proton energies above this value.

**4.3. $^{115}$In production**

The (p,α) reaction leads to the production of $^{115}$In in both the nearly stable ground state ($J^\pi = 9/2^+$, $T_{1/2} = 4.4 \times 10^{14}$ y) and the isomeric state $^{115m}$In ($J^\pi = 1/2^-$, $T_{1/2} = 4.486$ h). The isomeric state decays predominantly to the ground state (95%) and partly to stable $^{115}$Sn (5%). The independent production cross section of $^{115m}$In was determined using the IT decay γ-line at 336.24 keV.

## 4.4 $^{117}$Sn production

$^{117}$Sn can be produced via the (p,pn) and (p,d) channels in both the metastable state $^{117m}$Sn ($J^\pi = 11/2^-$, $T_{1/2} = 13.6$ d) and the stable ground state ($J^\pi = 1/2^+$). The metastable state decays entirely to the ground state through isomeric transition. The independent production cross section of $^{117m}$Sn was determined using γ-line at 158.56 keV. To ensure accurate measurements, gamma spectra were measured two days after irradiation, allowing sufficient time for the decay of interfering $^{117}$Sb (half-life = 2.8 hours), which also emits a γ-ray at the 158.56 keV energy. Consequently, the contribution of $^{117}$Sb became negligible, ensuring precise determination of the $^{117m}$Sn activity.

The reaction cross sections at different proton energies for all the reactions discussed are presented in Table 2. When calculating the uncertainties of the reaction cross sections, the following contributions were added in quadrature: statistical errors of the counts (0.5–22%), proton flux uncertainty (up to 7%), detector efficiency (up to 6.2%), and target thickness (5%). The resulting cross-section uncertainties lie in the range of 9.5–25%.

**Table 2.** Reaction cross sections at different proton energies. Energies ($E_{eff}$) were calculated using SRIM. The uncertainties include the SRIM-calculated standard deviation of the energy distribution, the contribution from the foil thickness, and uncertainties in the initial beam energy.

| Protons Energy (MeV) | Reaction cross-section (mb) | | | |
|---|---|---|---|---|
| | $^{118}$Sn(p,n)$^{118m}$Sb | $^{118}$Sn(p,2n)$^{117}$Sb | $^{118}$Sn(p,α)$^{115m}$In | $^{118}$Sn(p,x)$^{117m}$Sn |
| 17.53 ± 0.80 | 36.12 ± 4.11 | 843 ± 82 | 1.85 ± 0.2 | 90.33 ± 10.84 |
| 16.65 ± 0.64 | 41.87 ± 4.73 | 743 ± 79 | 1.65 ± 0.17 | 78.69 ± 9.4 |
| 15.86 ± 0.69 | 47.25 ± 5.11 | 654 ± 98 | 1.42 ± 0.15 | 59.53 ± 7.14 |
| 14.98 ± 0.68 | 60.62 ± 6.36 | 628 ± 93 | 1.36 ± 0.14 | 53.14 ± 6.38 |
| 14.07 ± 0.60 | 64.59 ± 6.51 | 464 ± 56 | 1.12 ± 0.12 | 32.58 ± 3.90 |
| 13.14 ± 0.61 | 78.17 ± 8.03 | 352 ± 79 | 0.96 ± 0.10 | 19.39 ± 2.33 |
| 12.17 ± 0.69 | 60.64 ± 6.36 | 133 ± 83 | 0.54 ± 0.06 | 3.76 ± 0.45 |
| 11.18 ± 0.81 | 53.55 ± 6.66 | – | 0.34 ± 0.04 | 2.69 ± 0.32 |
| 10.16 ± 0.83 | 37.79 ± 3.89 | – | 0.18 ± 0.02 | 1.052 ± 0.13 |
| 9.19 ± 0.79 | 25.87 ± 2.76 | – | – | 0.598 ± 0.07 |
| 8.11 ± 0.98 | 15.05 ± 1.65 | – | – | – |
| 6.86 ± 1.21 | 0.28 ± 0.03 | – | – | – |

## 4.5 Comparison with Theoretical Model Calculations (TENDL, JENDL)

In this section, we compare our experimental data with previously published results and with theoretical model calculations from the TENDL-2023 [33], TENDL-2025 [34] and JENDL [2] data libraries. Figures 6(a–d) present our experimental cross sections (listed in Table 2) together with previously published data [6, 25-26, 35]. It should be noted that in Refs. [25-26, 35] the measurements were performed using a $^{nat}$Sn target; therefore, for a consistent comparison with our results, their data were converted to correspond to a pure $^{118}$Sn target. In addition, evaluated data from the TENDL-2023, TENDL-2025, and JENDL-5 libraries are included.

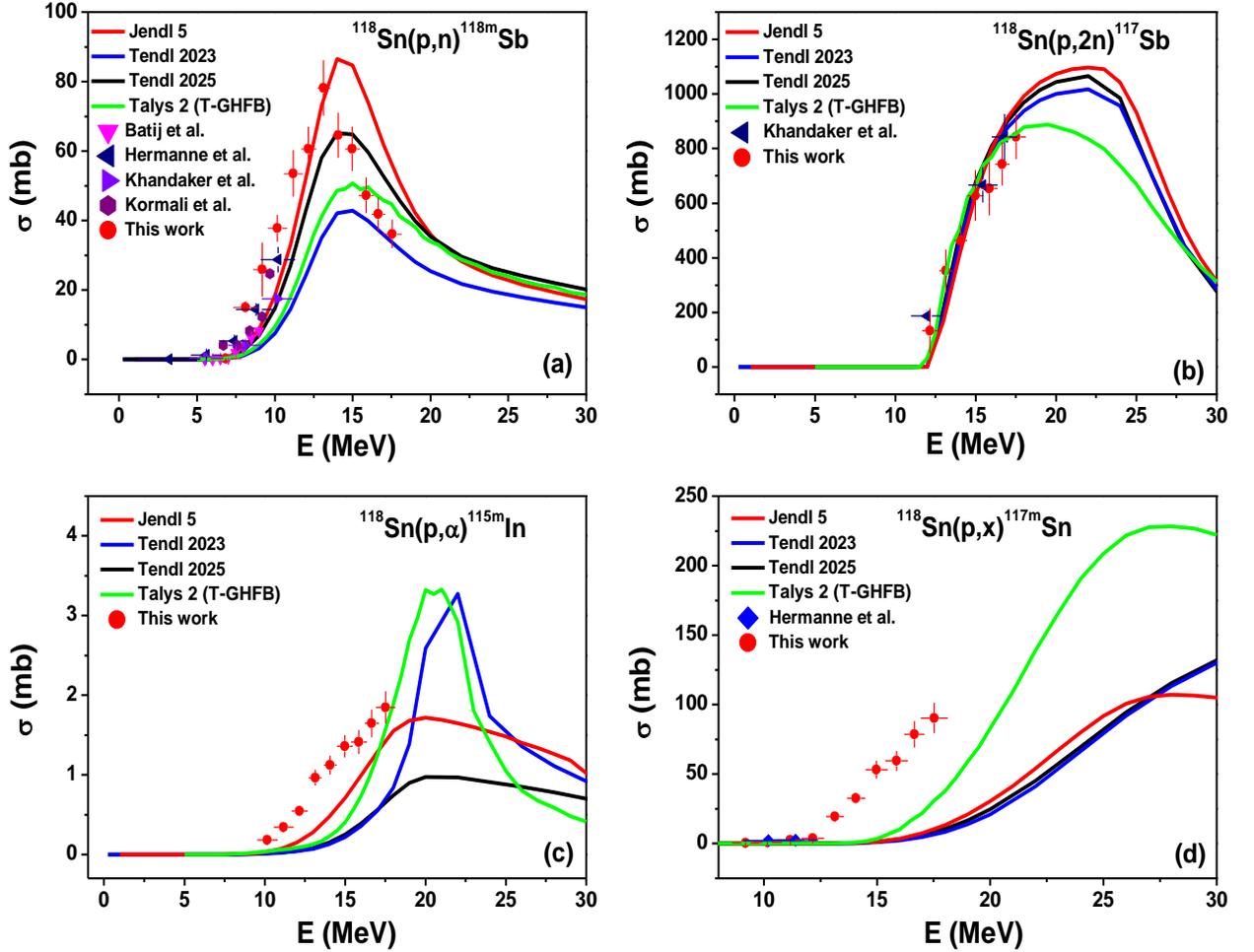

**Fig. 6** Excitation functions of the reactions: **(a)** $^{118}$Sn(p,n)$^{118m}$Sb, **(b)** $^{118}$Sn(p,2n)$^{117}$Sb, **(c)** $^{118}$Sn(p,α)$^{115m}$Sb, and **(d)** $^{118}$Sn(p,x)$^{117m}$Sb.

As seen from the figures, the experimental data are well-correlated, while the data from different theoretical databases show significant discrepancies, including the latest two versions of TALYS, namely TALYS-2025 and TALYS-2023 datasets. The only reaction in the considered energy range for which the theoretical calculations show results that are close to each other and to the experimental data is the $^{118}$Sn(p,2n)$^{117}$Sb reaction.

For the $^{118}$Sn(p,n) reaction leading to the metastable state $^{118m}$Sb, the three evaluated libraries yield notably different results. Among them, the JENDL-5 library demonstrates the best agreement with our experimental data, reasonably reproducing both the peak magnitude and the width of the excitation function (deviations do not exceed 29%).

The $^{118}$Sn(p,α) reaction leads to the formation of the metastable $^{115m}$In. It is well known that reactions involving the emission of composite particles, such as α-particles, are more difficult to describe theoretically compared to those involving the emission of one or two nucleons. In our previous work on the reactions $^{120}$Sn(p,α)$^{117m,g}$In [23,24], we observed significant discrepancies between TALYS predictions and experimental results. A similar trend is observed in this case: the TENDL-2023 and TENDL-2025 models do not reproduce the experimental data well, whereas JENDL-5 provides a much better description. However, even in this case, the average deviation between the calculated and experimental values remains around 66%.

The worst agreement between model predictions and experimental data was observed for the (p,pn) + (p,d) reaction, leading to the formation of the metastable $^{117m}$Sn (Fig. 6(d)). In this case, the excitation functions provided by all the evaluated databases are significantly shifted towards higher incident particle energies compared to our experimental results, suggesting potential deficiencies in modeling the primary reaction mechanisms for this channel.

In the analysis of reaction cross-sections at low and intermediate energies, three main components are typically distinguished: direct, pre-equilibrium, and compound processes, each of which can be analyzed using the corresponding theoretical models. The analysis of the compound component is usually performed using statistical theories based on nuclear level densities and transmission coefficients for incoming and outgoing particles.

Unlike JENDL-5, the TALYS code allows considerable flexibility in adjusting the parameters used in nuclear reaction calculations. Among these parameters, the choice of the nuclear level density (NLD) model is particularly important, since the calculated reaction cross sections are highly sensitive to the adopted level density description. The TALYS code includes several NLD models, which may lead to significantly different predictions for individual reaction channels.

In the present work, calculations were performed using TALYS with the temperature-dependent Gonyi–Hartree–Fock–Bogolyubov nuclear level density model [36]. Our previous studies [23,24] have shown that, among the level-density models available in TALYS, this approach provides the best overall agreement with experimental data for (p,α) reactions. Consistent with these earlier findings, the use of this level-density description leads to a somewhat improved reproduction of the experimental excitation functions for the $^{118}$Sn(p,α)$^{115m}$In and $^{118}$Sn(p,x)$^{117m}$Sb reactions (Fig. 6(c,d)). Nevertheless, even with this choice of the nuclear level density model, the agreement between calculated and experimental cross sections remains only partial.

In our view, a more important factor affecting the description of reaction channels involving the emission of composite particles may be related to possible cluster correlations in the nuclear structure [37]. Such correlations are expected to be strongly nucleus-dependent and therefore cannot be captured within a universal statistical description. In particular, experimental studies by Takayuki Tanaka et al. [38] reported indications of α-cluster structures in tin isotopes based on quasi-free (p,pα) knockout measurements, suggesting that α-like correlations may exist at the nuclear surface of Sn nuclei. These correlations can enhance the probability of composite-particle emission and thus influence the shape and position of the excitation functions. However, cluster degrees of freedom are not explicitly included in commonly used reaction codes such as TALYS, where reaction mechanisms are primarily treated within statistical and pre-equilibrium frameworks. As a consequence, the contribution of such effects may be underestimated in model calculations, which could partly explain the discrepancies observed between calculated and experimental excitation functions for reaction channels involving the emission of composite particles.

## 5 Conclusions

New experimental cross-section data have been obtained for proton-induced reactions on enriched $^{118}$Sn targets, specifically for the $^{118}$Sn(p,n)$^{118m}$Sb, $^{118}$Sn(p,2n)$^{117}$Sb, $^{118}$Sn(p,α)$^{115m}$In and

$^{118}$Sn(p,x)$^{117m}$Sn reactions. The excitation functions for these reactions were determined using the stacked-foil technique.

A detailed analysis of proton energy losses in the multilayer stack was carried out by comparing the average proton energies $E_{av}$ with the effective energies $E_{eff}$, the latter taking into account the proton energy distribution and the energy dependence of the reaction cross sections in each layer. The two approaches yield values differing by only 0.3–3%, confirming that the simple expression for $E_{av}$ provides an adequate approximation for stacks composed of nearly homogeneous foils with thicknesses of several tens of micrometers in the considered energy range.

The present results were compared with evaluated data from TENDL-2023, TENDL-2025, and JENDL-5. For the $^{118}$Sn(p,2n) channel leading to the ground state of $^{117}$Sb, the examined databases provide similar results within the considered energy range, which are in good agreement with the experimental data. In contrast, for reactions leading to metastable residual nuclei, the three libraries give significantly different results, with JENDL-5 showing better predictive ability compared to TENDL-2023 and TENDL-2025. For the (p,α) and (p,pn)+(p,d) channels, all model calculations are systematically shifted toward higher incident energies relative to experiment.

The largest deviations are observed for reaction channels involving the emission of composite particles, such as α particles and deuterons. One possible explanation is related to the presence of cluster correlations in the nuclear structure, which may enhance the probability of composite-particle emission. Since cluster degrees of freedom are not explicitly included in commonly used reaction codes such as TALYS, their contribution may be underestimated in model calculations, which could partly explain the observed discrepancies.

Overall, these findings indicate that while current evaluated nuclear data libraries offer a reasonable description of relatively simple reaction channels, further refinements are required to accurately account for more complex reaction mechanisms. In particular, processes involving composite-particle emission at lower proton energies warrant additional investigation to improve predictive capabilities of nuclear reaction models.

The research was supported by the Higher Education and Science Committee of MESCS RA (Research project 24WS-1C021).